\documentclass[useAMS,usenatbib]{aa}
\usepackage{txfonts}
\usepackage{natbib}
\usepackage{longtable}
\usepackage[dvips]{graphicx}
\usepackage{url}
\author{R. Voss\inst{1}, G. Nelemans\inst{1,2}}
\institute{$^{1}$Department of Astrophysics/IMAPP, Radboud 
University Nijmegen, PO Box 9010, NL-6500 GL Nijmegen, the Netherlands.\\
$^{2}$Institute for Astronomy, K.U. Leuven, Celestijnenlaan 200D, 3001 Leuven, 
Belgium.}
\title{Type Ia supernovae in globular clusters: observational
upper limits.}
\abstract
{}
{
In the dense stellar environment of the globular clusters,
compact binaries are produced dynamically. Therefore the fraction
of type Ia supernovae that explode in globular clusters is expected
to be higher than the fraction of mass residing in these. 
}
{
We have searched for globular clusters at the positions of observed
type Ia supernovae. We used archival HST images and literature data,
covering the positions either before the supernovae exploded, or
long enough after that the supernovae have faded below the luminosities
of globular clusters.
}
{
We did not find evidence for globular clusters at any of the
supernova positions. For 18 type Ia supernovae, the observations are
sensitive enough that any globular cluster would have been detected,
and for further 17 type Ia supernovae, the brighter globular clusters
would have been detected. Correcting for incompleteness, we derive
a 90\% upper limit of 0.09 on the fraction of type Ia supernovae that explode
in globular clusters for the full sample and 0.22 for the sample of
supernovae in late-type galaxies. This allows us to limit
enhancements per unit stellar mass for a coeval population
$\eta_{co}\lesssim50$ (100) with 90\% (99\%) confidence. We find that
by observing the positions of a sample of less than 100 type Ia 
supernovae in the outer parts of early-type galaxies, it will be possible 
to probe the currently favoured range of $\eta_{co}\sim1-10$.
}
{}
\begin{document}
\maketitle
\section{introduction}
Type Ia supernovae (SNIa) are believed to be thermonuclear explosions of
white dwarfs \citep[e.g.][]{Hillebrandt2000}. 
The two main scenarios for bringing the white dwarfs
above the critical explosion mass $M_C$ (similar but not equal to the
Chandrasekhar mass) are (1) the single-degenerate (SD) scenario, in which
a white dwarf accretes from a non-degenerate companion star 
\citep{Whelan1973,Nomoto1982}, and (2)
the double degenerate (DD) scenario, in which the merger of two white
dwarfs with a total mass exceeding $M_C$ causes the explosion 
\citep{Iben1984,Webbink1984}.
Both scenarios requires binaries with white dwarfs in orbits close
enough that mass transfer occurs at some point during the evolution
of the binary.

Globular clusters (GCs) are known to harbour large populations of compact
binaries. Most well known are the bright
low-mass X-ray binaries (LMXBs) with a specific density per unit
mass 100 times higher than in the field \citep{Clark1975}. 
Also the population of
millisecond pulsars \citep{Lyne1987,Lorimer2005} and blue stragglers 
\citep{Sandage1953} have been found to be 
strongly enhanced. Particularly interesting for type Ia supernovae
are the relatively recent observational evidence that also the populations
of white dwarf systems are enhanced 
\citep[e.g.][]{Heinke2003,Heinke2005,Dieball2007,Maccarone2007b,Knigge2008,Henze2009}. 
The enhancement of tight binaries in globular clusters is a consequence 
of the very high stellar densities found there 
(up to $10^{6}M_{\odot}$ pc$^{-3}$). With such high densities, dynamical
encounters which create or modify binaries are frequent, and the higher
mass of the binaries makes them sink to the center, where the 
encounter rates are highest. 

It is therefore reasonable to expect that both SD and DD SNIa progenitors
will be enhanced in globular clusters. The few theoretical studies
that have investigated this
have indeed found such enhancement in their models 
\citep{Ivanova2006,Shara2006}. However, the magnitude of such an
enhancement is highly uncertain. Aside from the fact that it is not
known which systems actually do lead to supernova explosions and how
these systems evolve outside globular clusters, there is
also a very large uncertainty due to the difficulties of modelling and
observing exotic binary populations in globular clusters. 

The most massive globular clusters, with the highest collision rates
are bright and can therefore be seen to large distances ($\sim$50-100 Mpc).
It is therefore feasible to put limits on the fraction of type Ia supernovae
that explode in globular clusters by identifying these in deep pre- or 
post-explosion images of the supernova positions \citep{Pfahl2009}. 
Despite this prediction, no such observational survey has been carried out. 

In this paper we perform a survey using archival observations
and literature data to place observational constraints on the
fraction of SNIae in globular clusters. We first discuss the theoretical
estimates in Sect. \ref{sect:theory}, then we discuss our methodology
in Sect. \ref{sect:method}. In Sect. \ref{sect:data}
we describe and
analyse the literature and archival data, and in Sect. \ref{sect:discuss}
we discuss the results and perspectives for future surveys.

\section{Theoretical expectations}
\label{sect:theory}
It is clear from both observations and theory that compact binaries
can be formed efficiently through dynamical interactions in the
dense stellar environment of globular clusters. However, the
picture of the formation and evolution of these compact
binaries remains vague, due to a large number of unconstrained
physical processes involved, and the computational expense
of globular cluster models including binary evolution.
Modelling the formation of SNIa progenitors is even more problematic,
given their unknown nature. It is therefore not surprising that only
few studies have attempted this.

The enhancement of compact binaries in a globular cluster is expected 
to be proportional to the rate of interactions $\Gamma$ 
\citep[e.g.][]{Hut1983,Pooley2003,Pooley2006}. While $\Gamma$
can be estimated for nearby globular clusters, based on their structural 
parameters, the estimates are not very reliable \citep{Maccarone2011}, 
and it is not possible
to measure the parameters accurately outside the Galaxy. For this
reason we use the commonly used {\it average} enhancement factor for a galactic
population of $N$ globular clusters
\begin{equation}
\label{eq:eta}
\eta=\frac{\sum\limits_{i=1}^{N} M_{GC,i}\times\eta_{GC,i}}{\sum\limits_{i=1}^{N}M_{GC,i}}.
\end{equation}

Single-degenerate progenitors were considered in the study of
\citet{Ivanova2006}, who found an enhancement factor (per unit
stellar mass) of $\eta$=1-7 compared to a field population with
solar metallicity. They find no single-degenerate SNIa in a population
of stars with the same properties as the globular clusters but with
interactions turned off. \citet{Shara2006} finds a small enhancement
$\lesssim$2 in the production rate of accreting white dwarfs, but find that
in their globular cluster models, the accreting white dwarfs are
heavier, and so more likely to be SNIa progenitors. Other studies
have predicted higher numbers of accreting white dwarfs
\citep{DiStefano1994,Davies1995}, corresponding to $\eta$=5, but
did not consider SNIa progenitors specifically.
\citet{Shara2006} find no enhancement of double white dwarf
binaries, but this study does not discuss SNIa progenitors
specifically. \citet{Ivanova2006} find an enhancement of merging double
white dwarfs above the Chandrasekhar mass of $\eta=$3-13 compared to
a field population with the same metallicity and age as the 
population in the globular clusters,
but no significant enhancement ($\eta=0.3-1.2$) when compared to 
field population with solar metallicity.

For the SD scenario, an alternative approach is to compare
to observations of similar systems. This is not possible for the
DD scenario, due to a lack of observational systems to compare to.
With their accreting massive white dwarfs, SD SNIa progenitors are 
somewhat similar to cataclysmic variables (CVs) and LMXBs in their 
formation and evolution. The bright LMXB population in globular clusters 
is quite well studied, as they can be seen to large distances with 
Chandra, and they have been found to be over-abundant by a factor
of $\sim$100 \citep{Clark1975,Sarazin2003,Jordan2007,Voss2009}. 
The CV population is much less understood, as they 
are much harder to identify. Only a small sample has been found in 
recent years, consistent with an over-production by a factor of 
$\sim$few \citep[e.g.][]{Pooley2006,Dieball2007,Knigge2008}, but completeness 
is a serious issue, and higher enhancement 
factors are therefore not ruled out. Observations of novae in
M31 suggests en enhancement factor of $\sim$10 \citep{Henze2009}.
The SD SNIa progenitors have white dwarfs with masses near $M_C$,
more similar to the masses of neutron stars than to those of most
CVs. They therefore sink the the center more easily and experience more 
dynamical encounters than more typical white dwarf systems, which
leads to a higher expected enhancement.

From the discussion above we conclude that $\eta$ is most likely 
greater than one and lower than 10. However, it is clear that the
results are very poorly constrained, and models with $\eta$ 
outside this range can not be discarded. It is
therefore important to constrain this fraction from observations.

Besides $\eta$, the fraction of SNIae that explodes in globular
clusters also depends on the fraction of stellar mass that
resides in the globular clusters, $F_{M,GC}=M_{GC}/M_{F}$, where
$M_{GC}$ is the total mass of the population of globular clusters
and $M_{F}$ is the total mass of all the other stars. For a sample of
$N$ SNIae, the expected number that explodes in globular clusters
is then:
\begin{equation}
\label{eq:N}
N_{GC}=N\times F_{M,GC} \times \eta.
\end{equation}
The mass fraction $F_{M,GC}$ varies strongly between galaxies.
The Milky Way has a low abundance of globular 
clusters, with $F_{M,GC}\sim0.1$\% \citep[e.g. the catalogue of][]{Harris1996}. 
Therefore only $\sim10$\% of the Milky Way LMXBs \citep[see e.g.][]{Liu2007},
and based on the considerations above, probably $\sim1$\% of the CVs, 
are in globular clusters. This
is in stark contrast to many elliptical galaxies with rich globular
cluster systems, where in some cases $F_{M,GC}$ can be larger than
1\% \citep[e.g.][]{Harris1999}, 
and the majority of bright LMXBs are in globular
clusters \citep{Angelini2001}. 
Correspondingly, if the SNIa enhancement is a factor of 10, 
about 10\% of the SNIae in these galaxies must be formed in globular 
clusters. Averaging over the population of nearby galaxies yields
a fraction of SNIae in globular clusters of $\sim1-3\cdot10^{-3}\eta$
\citep{Pfahl2009}, most likely a few per cent.

A complicating factor that was not taken into account in previous
studies is the fact that the rate of type Ia supernovae $R_{SNIa}$ 
decreases with age for a coeval population of stars.
In general the distribution
of globular cluster ages is different from that of the field stars.
Therefore 
\begin{equation}
\label{eq:rate}
\eta=\eta_{co}\times \frac{R_{SNIa}(t_{GC})}{R_{SNIa}(t_{f})}
\end{equation}
where $t_{GC}$ is the age of the globular clusters and $t_{f}$
is the age of the field stars. For early-type galaxies
$t_{GC}\sim t_f$, whereas the bulk of the field population of 
late-type galaxies tends to be significantly younger than the
globular cluster population. $\eta_{co}$ is the globular
cluster enhancement per unit stellar mass, compared to a
field population of {\it the same age}.
The exact shape of the delay-time distribution (DTD, the
SNIa rate as a function of time for a coeval population
of stars) is not known, but it has been shown to decrease
strongly by a factor of $>10$ from young environments 
$<$1 Gyr to older environments $\sim10$ Gyr \citep[e.g.][]{Maoz2010}.
Therefore $\sim$50-85\% of all type Ia supernovae explode within the first
Gyr after star formation \citep[e.g.][]{Maoz2010}. However, the
current star-formation density is much lower than at redshifts 1,
and comparing the local rate of star formation \citet{Hanish2006} 
to the stellar mass density \citet{Salucci1999,Cole2001} leads to
a fraction of stars in the local universe that were formed
less than 1 Gyr ago of $\sim2$\%. Despite the high fraction of
prompt SNIae for a \textit{coeval} population of stars, the
local population will therefore be dominated by the tardy component.
Combining the fractions found by \citet{Sullivan2006} with the local 
rate of star formation \citet{Hanish2006} and stellar mass density 
\citet{Salucci1999,Cole2001}, only $\lesssim20$ per cent of the local SNIae 
are expected to belong to the prompt component.

For most local galaxies, the ratio 
$R_{GC/F}=\frac{R_{SNIa}(t_{GC})}{R_{SNIa}(t_{f})}$
will therefore be higher than what could be expected from the DTD.
It is typically in the range $\sim0.1-1$. For early-type galaxies,
the typical age of the field population is similar to the age of
the globular clusters, and therefore $R_{GC/F}\sim1$. Late-type
galaxies can have significant populations of young stars for which
the SNIa rate is more than a magnitude higher than for the old
population of stars in their globular clusters. However, the vast
majority of late-type galaxies also have older stellar components,
and $R_{GC/F}$ will therefore almost always be higher than 0.1.

\section{Exclusion of cluster origins for a sample of type Ia supernovae}
\label{sect:method}
The association of type Ia supernovae with globular clusters relies
on the spatial coincidence. If a type Ia supernova is found to have
exploded at the same position as a globular cluster, it {\it may} have
exploded inside the globular cluster. If not, then a globular cluster
origin is definitely excluded. However, the distribution of GC
luminosities is wide, and many observations are only sensitive to
the bright end of the GC luminosity function. \citet{Pfahl2009}
showed that for $\eta$=10 approximately 1\% of all SNIa should explode
in globular clusters, corresponding to 1 supernova
in a globular cluster within 100 Mpc every year, and that a
dedicated HST program would be able to find the connection, if each
SNIa was observed a few years after the explosion.
With their assumptions such a program would require deep HST observations 
of $\sim$100 fields every year. They did not account for the fact that
some SNIa suffer from significant extinction and that observations
inside galaxies where there is a high background are much less sensitive
than in the field. These effects are difficult to model for the full
sample of type Ia supernovae, and it is therefore unclear if the proposed 
observational program would be succesful despite the high costs. 

We therefore use the currently available HST data to observe or put
limits on the fraction of SNIae in globular clusters, and to make it
possible to estimate the sensitivity that can be achieved with a dedicated
observational program. 

\subsection{Globular cluster completeness}
\label{sect:complete}
To be able to exclude a GC origin, it is necessary to know the
observable properties of the GCs. Being old stellar systems,
there are only relatively small variations in the mass-to-light
ratios of different clusters, with the main difference being
related to the globular cluster metallicity. However, the mass distribution
of globular clusters is wide, with several orders of magnitude difference
between the brightest and the faintest clusters 
\citep[e.g.][]{Harris1991,Jordan2007}. In the more
distant galaxies or in shallow observations it will
not be possible to observe the faint clusters and it is necessary
to calculate the incompleteness caused by this.

The most straight-forward way to do this is to use the observed
luminosity function of globular clusters \citep{Pfahl2009}.
However, while the definition of $\eta$ as the enhancement per
unit mass is a useful measure, it is misleading in terms of the
physical interpretation, as the number of compact binaries does
not scale linearly with the mass of the globular cluster.
Instead it is related to the stellar encounter rate $\Gamma$
inside the clusters. The distribution of structural parameters
(and hence $\Gamma$) of clusters is not well known. Since also
the exact processes of compact binary formation are poorly constrained,
it is therefore not possible to make theoretical estimates of
the relation between GC mass and the probability of hosting
SNIa progenitors.

We instead attempt to do this on an empirical basis. The only
compact binaries that have been surveyed in large samples of
GCs are LMXBs. We use the results of \citet{Peacock2010} and
\citet{Sivakoff2007} to estimate the
mass distribution of GCs that contribute to the population of compact
binaries, and thereby to estimate 4 completeness levels, CL25, CL50,
CL75 and CL100, meaning the GC masses/luminosities above which 25\%, 50\%,
75\% and 100\% of the compact binaries are expected to reside. We
strongly caution that this is based on observations of LMXBs
alone, and that it is very possible that SNIa progenitors could
have a different mass-dependence. However, currently there are
no models that predict a different behaviour, and the
populations of faint X-ray sources in Galactic GCs do seem to follow
the distribution of bright LMXBs.

To estimate the completeness masses, we use the K-band data of
\citet{Peacock2010} and the z-band data of \citet{Sivakoff2007}. 
We give the completeness values for
the K-band observations of M31 and z-band observations of Virgo in
table \ref{tab:CL}.
To compare these results and extrapolate them to other wavelengths
we use the integrated simple stellar population magnitudes of
\citet{Girardi2000,Marigo2008}. 
The values are in good agreement, assuming a 12 Gyr
stellar population with a Chabrier initial mass function and a
metallicity of 0.012. We also provide estimates of the corresponding
(initial) globular cluster masses, using the K-band magnitudes and 
two different metallicities. These are then used to find the colours in all the
different bands used in the following analysis. As the metal-rich
GCs are redder, they are fainter than the metal-poor GCs in the used bands, 
for a given K-band luminosity. We therefore use the calculations for
z=0.012 to determine the confidence limits in all the bands, noting
that in this way the confidence limits will be underestimated, decreasing
our sensitivity somewhat. Younger clusters would be brighter for
a given stellar mass, similarly leading to an underestimation of the
confidence limits. The magnitudes for CL100 are given in table
\ref{tab:CL100}. From table \ref{tab:CL} it can be seen that subtracting
1.5, 2.5 and 3.0 from these magnitudes yields CL75, CL50 and CL25,
respectively.

\begin{table}
\begin{center}
\caption{Empirical completeness levels (absolute magnitudes) of globular clusters with 
LMXBs in the K- and z-band, and the corresponding globular cluster masses (in $10^5M_{\odot}$)
from simple stellar population models.}
\label{tab:CL}
\begin{tabular}{lcccc}
\hline\hline
Completeness & CL100 & CL75 & CL50 & CL25\\
\hline
K-band & -9.5 & -11 & -12 & -12.5\\
z-band & -7.5 & -9.0 & -10.0 & -10.5\\
M z=0.012 & 1.7 & 6.7 & 17 & 27\\
M z=0.0012 & 2.5 & 9.9 & 25 & 40\\
\hline
\end{tabular}
\end{center}
\end{table}

\begin{table*}
\begin{center}
\caption{Assumed absolute magnitudes for CL100. The HST magnitudes are
for the VEGA photometric system. The differences between the magnitudes with
different instruments is too small to be of importance for our study.}
\label{tab:CL100}
\begin{tabular}{lccccccccccccccccccccc}
\hline\hline
Band & B & V & R & I & F555W & F569W & F675W & F814W & F850LP & F160W & F190N \\
\hline
Mag & -5.5   & -6.5   & -7.1   & -7.7   & -6.5   & -6.6   & -7.2   & -7.6   & -7.9 & -9.3 & -9.5 \\
\hline
\end{tabular}
\end{center}
\end{table*}

\section{Results}
\label{sect:data}
The observations of globular clusters at the positions of type
Ia supernovae can be performed before the explosions, but also
after, since the light of the globular clusters is provided by
stars, which are not affected by the explosion. As the luminosity
of the SNIae are very high just after the explosion, it is necessary
to wait until they have become fainter than a globular cluster.
This typically takes less than two years, although SNIae with
light-echoes might be bright enough to hide globular clusters
for a longer time.

\subsection{Literature survey}
\begin{table*}
\begin{center}
\caption{Faintest observed BVRI magnitudes or upper limits from late-time observations 
of type Ia supernovae in the literature. CL gives the completeness of
ruling out globular clusters at the position, see text.}
\label{tab:BVRI}
{\scriptsize
\begin{tabular}{lccccccccccc}
\hline\hline
Supernova & B & V & R & I & Galaxy & Gal type & Distance& $A_V$ &References& Type & CL\\ 
\hline
SN1990O & 22.747 & 22.886 & 23.629 & 22.368 &MCG+03-44-03& Sba & 150Mpc$^1$& $>0.29^{16}$ &8& & 0\\
SN1990N & - & 23.198 & - & - & NGC 4639 & Sb & 27.0Mpc$^1$& 0.221$^5$ &9 && 25\\
SN1991bg & - & 25.0 & - & -&NGC 4374 & E &18.5 Mpc$^2$ & 0.096$^5$ &10 & P (sublum) & 100\\ 
SN1992A & $>$26.5 & $>$26.5 & - & - &NGC 1380& S0& 21.2Mpc$^2$& 0.014$^5$& 10 && 100\\ 
SN1992bc & 22.716 & 22.190 & 23.172 & 22.935 &ESO 300-09& S& 101Mpc$^1$ & 0.012$^5$&12 && 0\\
SN1993L & - & 23.0 & -& 21.5 &IC 5270& Sc &23.7Mpc$^2$ & $>0.23^6$&10& & 0\\ 
SN1996X & 23.68 & 23.89 & 21.63 & 20.78 &NGC5061& E& 25.5Mpc$^2$& 0.031$^5$&11 && 25\\
SN1997cn & - & $>$23.198 &- &-&NGC5490 & E& 78.3Mpc$^3$& 0.0$^{12}$ &12 && 0\\
SN2000ce & 24.12 & 23.84 & 23.77 & 22.84 &UGC4195& SBc & 86.7Mpc$^1$& 1.67$^7$&13 && 0\\
SN2001C & 22.64 & 22.89 & 23.57 & 23.23 &LEDA 19975 & Sb& 47.6Mpc$^4$& 0.403$^{13}$&13&& 25\\
SN2001V & 22.60 & 21.96 & 23.09 & 21.84 &NGC 3987& Sb& 68.5Mpc$^1$ & 0.171$^5$& 13&& 0\\
SN2001bg & 21.99 & 22.10 & 22.61 & 21.61 &NGC 2608& SBb &36.3 Mpc$^2$& 0.868$^{13}$ &13&& 0 \\
SN2001dp & 20.40 & 21.49 & 21.38 & 19.62 &NGC 3953& SBb & 17.1Mpc$^2$& $0.09^{16}$&13 & & 25 \\
SN2003du & 22.771 & 22.827 & 23.010 & 22.121 &UGC 9391& SB & 44.5Mpc$^1$ & 0.032$^5$ & 14 && 0\\
SN2006gz & $>$24.4 & $>$24.2 & 25.5 & - &IC 1277 & Sc & 103Mpc$^4$& 0.753$^{15}$& 15 & P (overlum) & 0\\
\hline
\end{tabular}\\
References: $^1$\citet{Sandage2010}, $^2$\citet{Tully2009}, $^3$Average redshift-independent distance from NED, $^4$ redshift-based distance from NED, $^5$\citet{Hicken2009}, $^6$\citet{Cappellaro1997}, $^7$\citet{Krisciunas2001},$^{8}$\citet{Hamuy1996}, $^{9}$\citet{Lira1998}, $^{10}$\citet{Milne1999},$^{11}$\citet{Salvo2001}, $^{12}$\citet{Turatto1998}, $^{13}$\citet{Lair2006}, $^{14}$\citet{Stanishev2007}, $^{15}$\citet{Maeda2009}, $^{16}$\citet{Schlegel1998}.
}
\end{center}

\end{table*}
To facilitate comparisons and thereby the use of type Ia supernova 
as standard candles in cosmology, their lightcurves are observed
using BVRI photometry. While most studies are limited to less than
$\sim100$ days, a number of SNIae have published late-time data. 
We have surveyed the litterature to compile a sample
of 19 such SNIa. None of the published results show signs of a
constant component expected from a host cluster. We list these
in table \ref{tab:BVRI}. The table also lists the faintest observed
magnitudes in the BVRI bands, as well as the distance to the host
galaxy and an estimate of the extinction towards the supernovae.
We use these to find the intrinsic absolute magnitude of the faintest
observation in each band and compare these with the values given in
table \ref{tab:CL} and \ref{tab:CL100}, to find the confidence with
which we can exclude globular clusters on the basis of these
light-curves. In table \ref{tab:BVRI} we have left out four SNIae (SN1991T, SN2000E, SN2000cx
and SN2001el) for which useful late-time lightcurves exist, as more
constraining limits are found in the analysis below.

In table \ref{tab:Additional} we have compiled a second more
heterogenous literature sample. This consists of two old SNIae that
were observed before the definition of the BVRI photometry 
(SN1937C and SN1972E), one where the magnitude was reported using
HST with the VEGA magnitude system (SN2000cx) and 5 where upper
limits were published based on pre-supernova HST images. Similar
to table \ref{tab:BVRI} we derived completeness limits for these
supernovae.

\begin{table*}
\begin{center}
\caption{Additional published optical limits on SNIa in GCs, with indication of whether
the limit was from images obtained before or after the supernova explosion. HST magnitudes
are in the VEGA photometric system.
CL gives the completeness of ruling out globular clusters at the position, see text. }
\label{tab:Additional}
{\scriptsize
\begin{tabular}{lccccccccc}
\hline\hline
Supernova & When & Optical result & Galaxy & Gal type & Distance & $A_v$ &References & Type & CL\\ 
\hline
SN1937C & After &Magnitude 20 & IC 4182 & Sdm dwarf & 4.1Mpc$^2$ & $>0.04^{16}$&21 & & 0\\
SN1972E & After &Magnitude 17 & NGC 5253 & I0 & 3.4Mpc$^2$ & $>0.17^{16}$&22& & 0\\
SN2000cx & After &F555W=25.2 & NGC 524 & S0& 24.0Mpc$^2$& 0.248$^{31}$&23 & P(91T-like)& 75\\
SN2003cg & before &F814W=22.9 & NGC 3169 & Sb & 21.6 Mpc$^{3}$ & $>0.10^{16}$& 32 & &75\\
SN2003gs & before &F555W=25.1 & NGC 936 & SB0 & 23.0 Mpc$^{2}$ & 0.205$^{18}$ & 32&P(sublum, fast-decl) & 75\\
SN2003hv & before &F814W=25.7 & NGC 1201 & S0 & 20.2 Mpc$^{2}$ & 0.05$^{19}$ & 32& &100\\
SN2004W  & before &F850LP=26.0 & NGC 4649 & S0 & 16.5Mpc$^{3}$ & 2.4$^{30}$ &32&&100\\
SN2005df & before &F606W=26.8 & NGC 1559 & SBc & 15.4 Mpc$^{4}$ & $<0.40^{28}$ &32 &&100\\
SN2006dd & before &B=26.0,V=26.1,I=25.4 & NGC 1316 &Sa &17.8Mpc$^{20}$  &0.25$^{24}$ &24 &&100\\
SN2006mr & before &B=25.8,V=26,I=25.4  &  NGC 1316 & Sa &  17.8Mpc$^{20}$  & 0.25$^{24}$&24 && 100\\
SN2007sr & before &F814W=25, F555W=26.5 & NGC 4038 & Sc & 22.0Mpc$^2$  & 0.558$^{17}$&25 && 100\\
SN2007on & before &F475W=27 &  NGC 1404  & E & 20.2Mpc$^2$ & $>0.03^{16}$&26 && 100\\
SN2008ge & before &F606W=24.3 & NGC 1527    & S0& 18.0Mpc$^2$& 0.040$^{27}$ &27 & P(02cx-like)& 75\\
SN2011fe & before &F814W=26.35 & M81 & Sb & 6.4 Mpc$^{28}$ & $>0.04^{16}$& 29 & & 100\\
\hline
\end{tabular}\\
References: 1-16 see table \ref{tab:BVRI}, $^{17}$\citet{Milne2010},$^{18}$\citet{Krisciunas2009},$^{19}$\citet{Leloudas2009},$^{20}$\citet{Stritzinger2010},$^{21}$\citet{Baade1956}, $^{22}$\citet{Kirshner1975}, $^{23}$\citet{Sollerman2004}, $^{24}$\citet{Maoz2008}, $^{25}$\citet{Nelemans2008}, $^{26}$\citet{Voss2008},
$^{27}$\citet{Foley2010}, $^{28}$\citet{Shappee2011}, $^{29}$\citet{Li2011}, $^{30}$\citet{Elias2006},$^{31}$\citet{Li2001},
$^{32}$S.J. Smartt, private communication, see \url{http://www.lorentzcenter.nl/lc/web/2010/391/presentations/Smartt.ppt}.
}
\end{center}
\end{table*}

\subsection{HST observations}
\label{sect:HST}
In addition to the literature survey, we have searched for
archival HST data at the positions of all known SNIa within 100 Mpc.
We analysed the data to either find observations where the supernova is faint
enough to exclude a GC origin, or where the supernova was not observed
and an interesting upper limit could be inferred. One problem is
the positional accuracy of the sources, as the fields of external
galaxies can be crowded. Therefore positions of sub-arcsec precision
are needed. Unfortunately many supernova positions are relatively poorly 
determined, as the only published coordinates are from the original
detections with small telescopes, when the supernovae are very bright,
and the coordinates are provided without error estimates.
In the case where several groups have published coordinates, the
distance between the positions can be large. Unfortunately most supernova 
observations are not publicly available, making it impossible to verify the positions.

We therefore only report results from supernovae, where we are confident 
that the positions are well-determined. In a few cases, the limits were
obtained from images, where the supernova is still seen at a luminosity
below that of globular clusters. In some cases, the
positions could be found from other HST images taken when the supernova
was visible, which were then matched to the image from which the limit
was determined. In the cases where the positions could not be determined 
from HST images, we used the positions listed in \citet{Hicken2009}, who
performed careful astrometry of a large sample of supernovae. For each
of the supernovae where we used the \citet{Hicken2009} positions we compared
to HLA HST images, where the HST astrometry was already corrected using
either 2MASS, GSC or SDSS as reference. We also used these catalogues
to confirm the astrometry, as the HLA astrometry is done automatically.    
We furthermore searched for ground-based observations of the supernovae,
which allowed us to find the positions of 7 additional SNIae.

To find the upper limit in an individual observation, we used the
background counts in a number of circular regions near the supernova
position (excluding point sources) to estimate the count rate of a 
3-sigma fluctuation $C_{3\sigma}$. The upper limit $C_{UL}$ on the supernova 
counts were then calculated by subtracting the median $C_\mu$ of
the background fields $C_{UL}=C_{3\sigma}-C_\mu$. The radius of the 
circular regions were chosen
in the range 0.15-0.5 arcsec, depending on the local density of point
sources and the gradient of the host galaxy light. From this we calculated
the upper limit in the VEGA magnitude system using the standard count-rate to
magnitude conversions from the HST data handbook, including PSF corrections
for the aperture size. We tested our method against the method used in
\citet{Nelemans2008,Voss2008}, where fake sources were put in the images
and the upper limit was based on the detection of these, and we found
good agreement.

The results are listed in table \ref{tab:HST}, for the supernovae where
the upper limits provide constraints on the globular cluster connection.

\subsubsection{Notes on individual sources.}
Some of the analysis of individual sources needs more explanation. SN1998bu
and SN1991T have light echoes that are seen in all the HST observations. 
For these sources we use the last F814W observations, and we measure
their flux within an aperture of 0.2 arcsec radius from the central source
position. For both SNIae, the flux from this region is low enough that
the existence of globular clusters at these positions can be excluded.

For many of the SNIae in table \ref{tab:HST} the position can not be
found from HST data. Four positions (SN2002bo, SN2002fk, SN2005el and SN2005hk)
are from \citet{Hicken2009}, and further 7 were found from ground-based
data that we matched to the HST images. Four 
SN1994ae, SN2001el, SN2005df and SN2007af were identified in
images taken with the NTT (SN1994ae) and VLT (SN2001el, SN2005df and SN2007af) 
obtained from the ESO archive. 
Three more SNIae positions were found from ground-based images
provided by Weidong Li. SN1999gh and SN1999gd were observed with
KAIT and SN1998aq with a 1.2m CfA telescope.

\begin{table*}
{\scriptsize
\begin{center}
\caption{SNIae for which we were able to exlude a globular cluster origin based on
archival HST observations. The time indicates the year of the observation used for the
exclusion. The VEGA photometric system was used for the apparent magnitudes. 
CL gives the completeness of
ruling out globular clusters at the position, see text.}
\label{tab:HST}
\begin{tabular}{lccccccccccc}
\hline\hline
Supernova & Instrument &Band & Time & Observation name & Magnitude & Galaxy &  Gal type &Distance& $A_V$& Type & CL\\
\hline
SN1991T & ACS HRC & F814W & 2006 & 10607 01 & 24.4 & NGC 4527& Sb & 17.6Mpc$^2$ & 0.302$^5$ & P (overlum) & 100\\
SN1994aa & WFPC2 & F791W & 1997 & 06419 01 & 23.2 & NGC 1320 & Sa & 37.7 Mpc$^{34}$& $>0.2^{16}$& & 100\\
SN1994ae & WFPC2 & F814W & 2001 & 09042 41 & 24.8 & NGC 3370 & E & 27.4 Mpc$^3$ &0.09$^{35}$&&100\\
SN1994D & ACS WFC & F850LP & 2003 & 9401 08 & 21.8 & NGC 4526& S0 & 16.9Mpc$^2$& 0.009$^5$ && 50\\
SN1998aq & ACS WFC & F555W & 2007 & 10802 8f & 23.6 & NGC 3982 & S & 22.0Mpc$^3$ & 0.04$^{35}$ && 50\\
SN1998bu & ACS HRC & F814W & 2006 & 10607 02 & 25.3 & NGC 3368 & Sab & 7.2Mpc$^2$ & 0.631$^5$ && 100\\
SN1999by & ACS WFC & F814W & 2004 & 9788 13 & 25.5 & NGC 2841 & Sb & 14.4Mpc$^2$ &0.030$^5$ & P (sublum) & 100\\
SN1999gd & ACS WFC & F814W & 2004 & 9735 32 & 26.1& NGC 2623 & Sb & 92.4Mpc$^3$ & 0.13$^{35}$&&75\\
SN1999gh & WFPC2 & F702W & 1999 & 06357 14 & 24.3 & NGC 2986 & E & 35.3Mpc$^3$& 0.18$^{35}$ && 75\\
SN2000E & ACS WFC & F814W & 2003 & 9788 d9 & 25.1 & NGC 6951 & SBb & 23.0Mpc$^2$&0.466$^5$ & &100\\
SN2001el & ACS HRC & F435W & 2006 & 10883 16 & 25.3 & NGC 1448 & Sc & 22.2Mpc$^{1}$ & 0.500$^{5}$ & &50\\
SN2002fk & ACS WFC & F814W & 2005 & 10497 01 & 23.5 & NGC 1309 & Sc & 24.1Mpc$^2$ & 0.034$^5$ & &75\\
SN2005cf & WFPC2 & F814W & 2007 & 10877 18 & 24.3 & MCG-01-39-03 & S0 & 29.9Mpc$^1$ & 0.208$^5$ && 75\\
SN2005el & WFPC2 & F814W & 2008 & 10877 66 & 24.3 & NGC 1819 & SB0 &69.5Mpc$^1$ & 0.012$^5$ && 50\\
SN2005hk & NIC2 & F160W & 2008 & na1p04010 & 24.4 & UGC 272 & Sc & 23.0Mpc$^2$ & 0.810$^5$ & P (02cx-like) & 100\\
SN2006X & WFPC2 & F814W & 1996 & 06584 02 & 23.6 & NGC 4321 & Sc & 13.2Mpc$^2$ & 2.496$^5$ &  & 50\\
SN2007af & WFC3 & F160W & 2010 & ib1f42010 & 22.7 & NGC 5584 & Sc & 22.6Mpc$^{3}$& 0.12$^{33}$ & &100\\
\hline
\end{tabular}\\
References: 1-16 see table \ref{tab:BVRI}, $^{33}$\citet{Brown2010},$^{34}$\citet{Tully1988}, $^{35}$\citet{Jha2006}.
\end{center}
}
\end{table*}


\subsection{Derived upper limits}
The results above constitute the first observational survey of the 
connection between globular clusters and type Ia supernovae. 
From this it has been possible to 
exclude a connection fully or partially for 35 SNIae, and no SNIa with possible 
globular counterparts has been found.
Taking into account that faint globular cluster counterparts would still be 
possible for some of the SNIae in our sample, this allows us to derive 
$P_{CL}=90$\% and 99\% upper limits on the fraction of SNIae in 
globular clusters $F_{UL}$. If a fraction $F_{SNIa}$ of SNIae explode
in globular clusters, the probability of {\it not} finding a globular cluster
at the position of SNIa number $i$ is $1-F_{SNIa}\times CL_i$, where
$CL_i$ is the globular cluster completeness level (which for our
sample can take the values 0.25, 0.5, 0.75 and 1.0, see section
\ref{sect:complete}). The probability of finding no globular 
clusters for the entire sample of $N$ SNIae is the product of finding
no globular clusters for each SNIae and $F_{UL}$, the upper limit on
$F_{SNIa}$, can therefore be found by solving
\begin{equation}
\prod\limits_{i=1}^N (1-F_{UL}\times CL_{i})=1-P_{CL}.
\end{equation}

Table \ref{tab:limits} lists the limits for the full
sample, and for two subsamples. The first subsample consists of all the
{\it normal} SNIae, which follow the same standard lightcurve evolution.
This sample is chosen to eliminate the ones with peculiar lightcurves,
as they are more likely to have been studied in detail
and therefore can bias our sample by being over-represented. Furthermore, 
their origin might differ from the normal SNIae. We note that the peculiar
label is subjective and some of them could belong to the normal population,
making the normal sample overly conservative.
Our second subsample consists of all the SNIae in early-type (elliptical and S0) galaxies. 
These have old stellar populations that are similar to the populations found in globular 
clusters, and are therefore particularly useful for comparison.

The 90\% upper limit from the full sample is very close to our expectation that
a few per cent of type Ia supernovae could explode in globular clusters, and it is
therefore useful for constraining theoretical models. From equations \ref{eq:N} and
\ref{eq:rate}, the upper limits on $F_{UL}$ corresponds to upper limits on
$\eta$:
\begin{equation}
\eta_{UL}=\frac{F_{UL}}{F_{M,GC}\times<R_{GC/F}>}
\end{equation}
For the full sample, the average stellar population is significantly
younger than in the globular clusters, and we therefore assume
$<R_{GC/F}>\sim0.2-0.4$. As the fraction of globular clusters is small
in the late-type galaxy part of the sample, $F_{M,GC}\sim0.1-0.3$\%. 
With these assumption we get a
result of $\eta_{co}<75-450$.
For the sample of early-type galaxies, the ages of the stars are similar to the
ages of the globular clusters and therefore $<R_{GC/F}>\sim1$.
Furthermore, the globular cluster mass fraction is higher, and we assume 
an average
value of $0.3-0.6$\%. This leads to a more constraining limit of $\eta_{co}<37-73$.

\section{Discussion}
\label{sect:discuss}
We have derived upper limits on the enhancement
per unit stellar mass $\eta_{co}$ of SNIae in globular clusters.
The limits are above the favoured theoretical expectations by almost
an order of magnitude, but are well below the observed enhancement of LMXBs,
$\eta_{LMXB}>100$. We have discussed the
effect of the decline of the SNIa DTD on the observations. Due to
the older age of globular clusters compared to the field population
in late type galaxies, this decreases the value of the parameter
$\eta$ (which can be seen as the average enhancement over the whole
population of galaxies), which is usually used in the literature 
(see equation \ref{eq:rate}). For this reason we find that the
sample of SNIae in early-type galaxies is more constraining that
the full sample, despite containing only $\lesssim$1/3 of the supernovae.
As this effect of the DTD will be sample-dependent
we believe that it is more appropriate to use $\eta_{co}$,
which is the direct measure of the enhancement factor. However, we
note that it is possible that also $\eta_{co}$ is time-dependent,
as early and late SNIae progenitors are likely to have different 
evolutionary histories, and since the structure of globular clusters
evolve.

In this pilot study we have only included supernovae which had HST 
images available, and for which we were able to identify the position 
with high accuracy.
This sample constitutes only a fraction of the total number of SNIae
within distances at which it is possible to observe globular clusters
($\sim$100 Mpc). As no globular clusters were detected at the position
of the supernovae, and as we have managed to probe enough SNIae positions
that we are starting to constrain theory, it is clearly interesting to
extend the sample. To reach the values currently favoured by theory,
it will be necessary to expand the sample by a factor of few.

There are several ways to proceed. As new supernovae
are discovered continuously, and more galaxies are being observed with
HST, it is possible to simply wait and let the sample expand. However,
it is clear that from the current sample that only a few SNIae will be
added every year, and it will therefore take decades before we can
reach the goal of a factor of few more supernovae. Another option is
to use archival data from other telescopes. As none of these are as 
sensitive to globular clusters as HST, it will only be possible to
probe more nearby SNIae, but the advantage is that many more galaxies
will be covered by useful observations. Such a study will be tedious,
as data from many different telescopes will have to be processed,
and it is unlikely that a factor of few can be reached.

A third option is to make a dedicated obsertional program to
observe the positions of known SNIae. Enough are known that
it will be possible to reach the goal of increasing the sample
by a factor of few. The most effective way of reaching it would
be to make a joint program with a medium-sized ground-based
telescope for the nearest SNIae, and HST for the more distant
ones. Our experience from this pilot project tells us that it
is very difficult to obtain useful limits for sources well
inside the galaxies as the background there is bright and
inhomogenous. While the number of SNIae will be lower if one
excludes sources in the inner parts of the galaxies, it will
be possible to be complete to larger distances, $\sim25-30$ Mpc
with ground-based telescopes and 100 Mpc with HST.

From our results it can clearly be seen that targetting early-type
galaxies will provide much stronger constraints due to the
lower difference between the ages (and hence SNIa rate) of the
field and globular cluster populations, and this effect is being
enhanced by the higher incidence of globular clusters in these
galaxies. By specifically targetting SNIae at larger radii, can
further reduce the number of observations needed. This is because
the globular clusters have shallower radial density profiles than
the stars in the galaxies. For this reason, $F_{M,GC}$ can be
several times higher when only considering the outer parts of
galaxies. It can therefore be possible to probe values of $\eta_{co}\sim$few
with less than 100 positions.

\section{Conclusions}
We have searched for globular clusters at the position of observed
type Ia supernovae, using archival HST observations and literature
data. We did not find evidence for globular clusters at any of the
SNIa positions, and our analysis showed that for 18 SNIae the
observations were sensitive enough that any globular cluster
should have been detected if it was there. For the positions of
further 17 SNIae, bright globular clusters would have been detected.
For the latter sample, we have developed a new empirical method to 
estimate the incompleteness, based on samples of globular clusters with
bright X-ray sources in nearby galaxies. The sample of non-detections
allow us to derive upper limits on the fraction of type Ia supernovae
that take place in globular clusters. For the full sample the
90\% and 99\% upper limits on the fraction of type Ia supernovae
are 9\% and 16\%. This is higher than
the currently favoured estimates of $\lesssim1$\%, but within the range allowed
by theory. The sample of early-type galaxies provides the best limits
for the enhancement factor per unit stellar mass for a coeval population
$\eta_{co}\lesssim$50 (90\% confidence).
We argue that a dedicated survey, combining ground-based
observations of nearby SNIa positions with HST observations of more distant
SNIa positions would be able to probe the favoured theoretical estimates.

\begin{table*}
\caption{Upper limits on the fraction of type Ia supernovae in globular clusters. The
table lists the sum of the completeness (CL) for each SNIa in the sample. $F_{UL,90\%}$
and $F_{UL,99\%}$ are the upper limits on the fraction of SNIae in globular clusters
derived from the sample, $<R_{GC/F}>$ and $\eta_{co,90\%}$ are the ranges of assumed values of
the fraction of stellar mass in the globular clusters and the SNIa rate ratio for the
age of the globular clusters compared to the field, respectively. $\eta_{co,90\%}$ and
$\eta_{co,90\%}$ are the upper limits on the values on the coeval enhancement of the
SNIa rate in globular clusters per unit stellar mass, $\eta_{co}$.}
\label{tab:limits}
\begin{tabular}{lllllllllll}
\hline
\hline
Sample & Sum & $F_{UL,90\%}$ & $F_{UL,99\%}$ & $F_{M,GC}$ & $<R_{GC/F}>$ & $\eta_{co,90\%}$ &$\eta_{co,99\%}$\\
\hline
Full sample & 27.5 & 9\%& 16\% & 0.3-0.1\% & 0.4-0.2 & 75-450 & 133-800\\
Normal SNIae & 21.25 & 11\%& 20\% & 0.3-0.1\% & 0.4-0.2 & 92-550 & 166-1000\\
SNIae E+S0 & 9.75 & 22\%& 39\% & 0.6-0.3\% & 1 & 37-73 & 65-130\\
Normal SNIae E+S0 & 7.25 & 28\% & 49\% & 0.6-0.3\% & 1 & 46-94 & 82-163\\
\hline
\end{tabular}
\end{table*}

\begin{acknowledgements}
This research is supported by NWO Vidi grant 016.093.305.
We thank Weidong Li for help with finding supernova images.
Based on observations made with the NASA/ESA Hubble Space Telescope, 
obtained from the data archive at the Space Telescope Institute. STScI 
is operated by the association of Universities for Research in Astronomy, 
Inc. under the NASA contract  NAS 5-26555.

\end{acknowledgements}

\bibliographystyle{/home/voss/work/bibliography/aa}
\bibliography{/home/voss/work/bibliography/general}

\end{document}